\documentclass{INTERSPEECH2023}
\interspeechcameraready
\usepackage{xcolor}
\usepackage{comment}
\usepackage{multirow}
\usepackage{siunitx}
\usepackage[acronym]{glossaries}
\usepackage{subcaption}

\makeglossaries
\newacronym{asr}{ASR}{Automatic Speech Recognition}
\newacronym{CNN}{CNN}{Convolutional Neural Network}
\newacronym{LSTM}{LSTM}{Long Short Term Memory}
\newacronym{tLSTM}{T-LSTM}{time LSTM}
\newacronym{fLSTM}{F-LSTM}{frequency LSTM}
\newacronym{mvfLSTM}{mvF-LSTM}{multiview frequency LSTM}
\newacronym{lstft}{LSTFT}{Log Short Time Fourier Transform}
\newacronym{lfbe}{LFBE}{Log FilterBank Energy}
\newacronym{lfr}{LFR}{Low Frame Rate}
\newacronym{rfr}{RFR}{Regular Frame Rate}
\newacronym{LAS}{LAS}{Listen Attend and Spell}
\newacronym{SNR}{SNR}{signal to noise ratio}
\newacronym{WER}{WER}{word error rate}
\newacronym{MHSA}{MHSA}{multihead self-attention}
\newacronym{rWERR}{rWERR}{relative word error rate reduction}
\newacronym{MLE}{MLE}{maximum likelihood estimation}

\title{Multi-View Frequency-Attention Alternative to CNN Frontends \\ for Automatic Speech Recognition}
\name{Belen Alastruey$^1$, Lukas Drude$^2$, Jahn Heymann$^2$, Simon Wiesler$^2$}
\address{
  $^1$TALP Research Center, Universitat Politècnica de Catalunya, Barcelona, Spain\\
  $^2$Amazon Alexa, Germany}
\email{belen.alastruey@upc.edu, \{drude, jahheyma, wiesler\}@amazon.com}

\begin{document}

\maketitle
\begin{abstract}
Convolutional frontends are a typical choice for Transformer-based \gls{asr} to preprocess the spectrogram, reduce its sequence length, and combine local information in time and frequency similarly.
However, the width and height of an audio spectrogram denote different information, e.g., due to reverberation as well as the articulatory system, the time axis has a clear left-to-right dependency.
On the contrary, vowels and consonants demonstrate very different patterns and occupy almost disjoint frequency ranges.
Therefore, we hypothesize, global attention over frequencies is beneficial over local convolution.
We obtain \SI{2.4}{\percent} \gls{rWERR} on a production scale Conformer transducer replacing its \gls{CNN} frontend by the proposed F-Attention module on Alexa traffic.
To demonstrate generalizability, we validate this on public LibriSpeech data with an \gls{LSTM}-based \gls{LAS} architecture obtaining \SI{4.6}{\percent} \gls{rWERR} and demonstrate robustness to (simulated) noisy conditions.
\end{abstract}

\section{Introduction}
The advent of Transformer-based models \cite{transformer} has surpassed the barriers of text. In \gls{asr} the Transformer typically operates on audio features like the mel-spectrogram \cite{dong_icassp}. These features provide longer input sequences compared to their text counterparts, and benefit from a frontend before the Transformer-based encoder \cite{gulati20_interspeech, gong21b_interspeech}.
This frontend preprocesses the features and reduces their sequence length.
The frontend of choice is often a \gls{CNN} \cite{gulati20_interspeech, wang-etal-2020-fairseq}, employing a stride for the sequence length compression.
However, these layers were originally designed for image processing \cite{cnns_lecun}, and although direct use has been shown to work well, it might be sub-optimal due to the following reasons.
Time and frequency are handled the same way although they contain different information.
Furthermore, although speech information is usually concentrated in lower frequencies, the model is forced to process all frequency ranges equally. And finally, since attention is used just on tokens obtained after splitting the features time-wise, the extraction of frequency interactions is limited by the usually small kernel size in the \gls{CNN} frontend.
\begin{figure}[t]
    \centering
    \includegraphics[scale=.15]{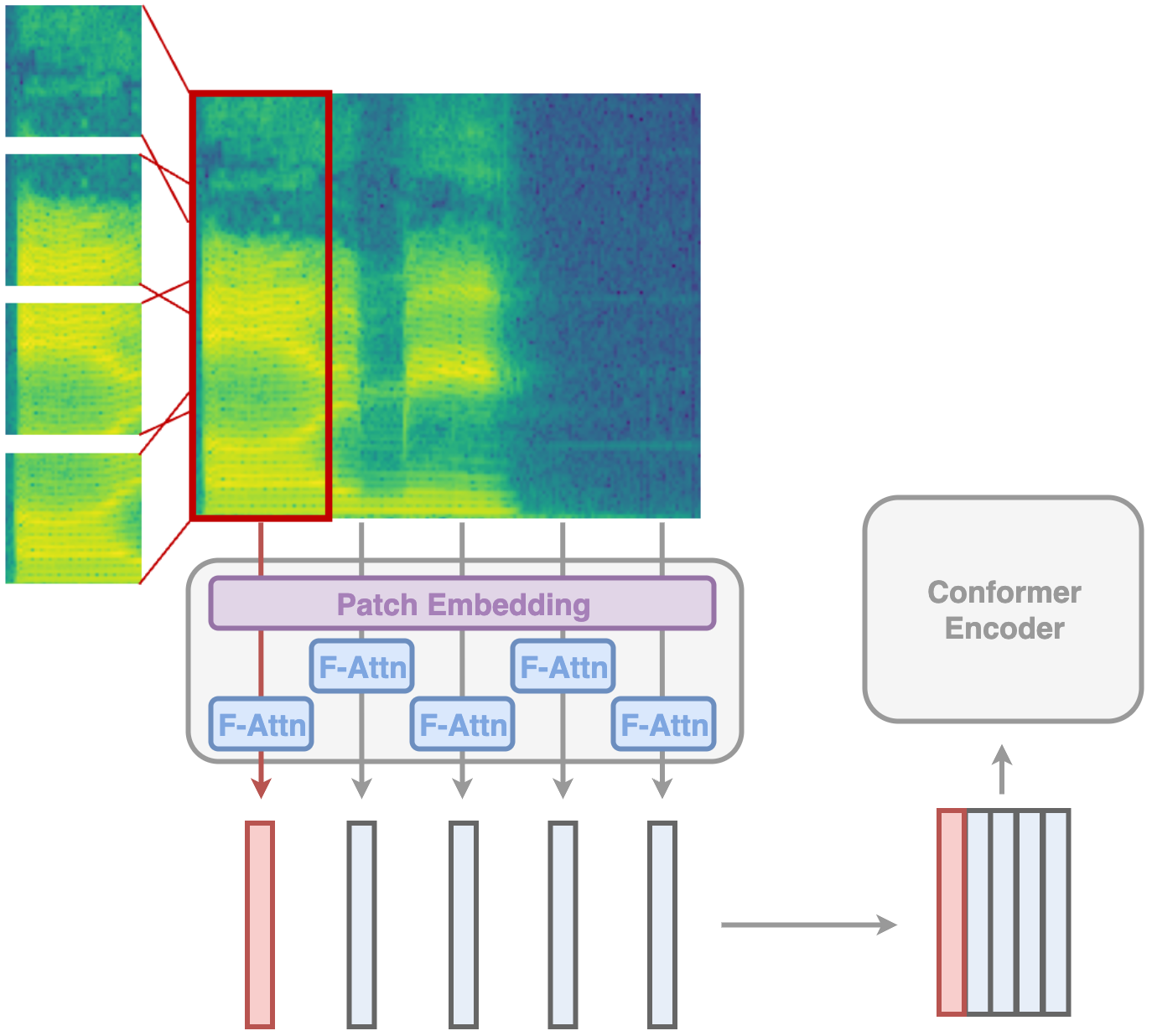}
    \caption{Overview of the proposed F-Attention frontend. The spectrogram is split into overlapping patches. Then, we use attention to extract interactions between patches on the same time range but on different frequency bins. We concatenate the outputs time-wise to obtain the input of the Conformer encoder.}
    \label{fig:fattention_frontend}
\end{figure}

In the past, \glspl{fLSTM} have been proposed as a frontend to \gls{LSTM}-based acoustic models to better capture frequency dependencies.
The frontend operates on time steps independently and scans the different frequency bins.
Then, the hidden states of the \gls{fLSTM} are concatenated and used as input for the acoustic model.
An extension of this method, the \gls{mvfLSTM} \cite{mvflstm}, uses many \glspl{fLSTM} with different window sizes and strides (referred to as "views") for each time step.
Then, the output of each \gls{fLSTM} is combined using a projection layer and the resulting vector is the input of the acoustic model.

Although \gls{fLSTM} could seem an adequate alternative to solve the issues of \gls{CNN} frontends, these networks are not suitable either to preprocess a spectrogram when using a Transformer-based model.
The concatenation of the different hidden states of the \gls{fLSTM} on each time step causes an increase of the embedding dimension by a factor of 8.
As a consequence, the number of parameters in the projection matrices in the attention layers of the Transformer-based encoder would increase remarkably.
\Glspl{mvfLSTM} bypass this problem through the use of a projection layer that is intended to merge the different views, and that can simultaneously be used to compress the embedding dimension. However, the number of parameters on this linear layer scales drastically when adding views, and would not be comparable to a \gls{CNN} frontend. But most importantly, neither \glspl{fLSTM} nor \glspl{mvfLSTM} compress the sequence length. This can be a problem when regarding complexity, since the Transformer’s attention matrix computational cost scales with $\mathcal O(n^2)$, where $n$ is the sequence length. 

Other frontend alternatives have been inspired by vision Transformers \cite{dosovitskiy2021an}, such as the Audio Spectrogram Transformer for audio classification \cite{gong21b_interspeech}. Then, the patches are embedded and reordered to form a sequence that is used as the input of a BERT-like classifier \cite{devlin-etal-2019-bert}. Thanks to the patches rearrangement for the input sequence formation, this frontend allows the Transformer-based encoder to analyze the interaction between different frequencies (patches corresponding to the same time range but to different frequency bins will be in separate tokens in the final sequence). However, this approach does not reduce the sequence length either, and in fact, the final sequence is even longer than the original one. Furthermore, the sequence obtained after the patches rearrangement is interleaved which may be problematic for a sequence-to-sequence task. 

In this work, we present a new frontend based on two main ideas: (1) patch extraction, inspired by vision Transformers, and (2) the analysis of frequency interactions, inspired by \glspl{fLSTM} and \glspl{mvfLSTM}. We propose the F-Attention frontend, which extracts patches out of a spectrogram and then uses attention to extract the relationships between patches on a same time range but on different frequency bins.
This frontend aims to improve the way speech features are processed, by being able to focus on relevant frequency ranges. Our proposed frontend obtains performance improvements with respect to the \gls{CNN} baseline. We show how F-Attention is able to dismiss noise and focus on relevant frequency ranges using an interpretability method and evaluate noise robustness.

\section{mvF-Attention Frontend}
\begin{figure}[h]
    \centering
    \includegraphics[scale=.065]{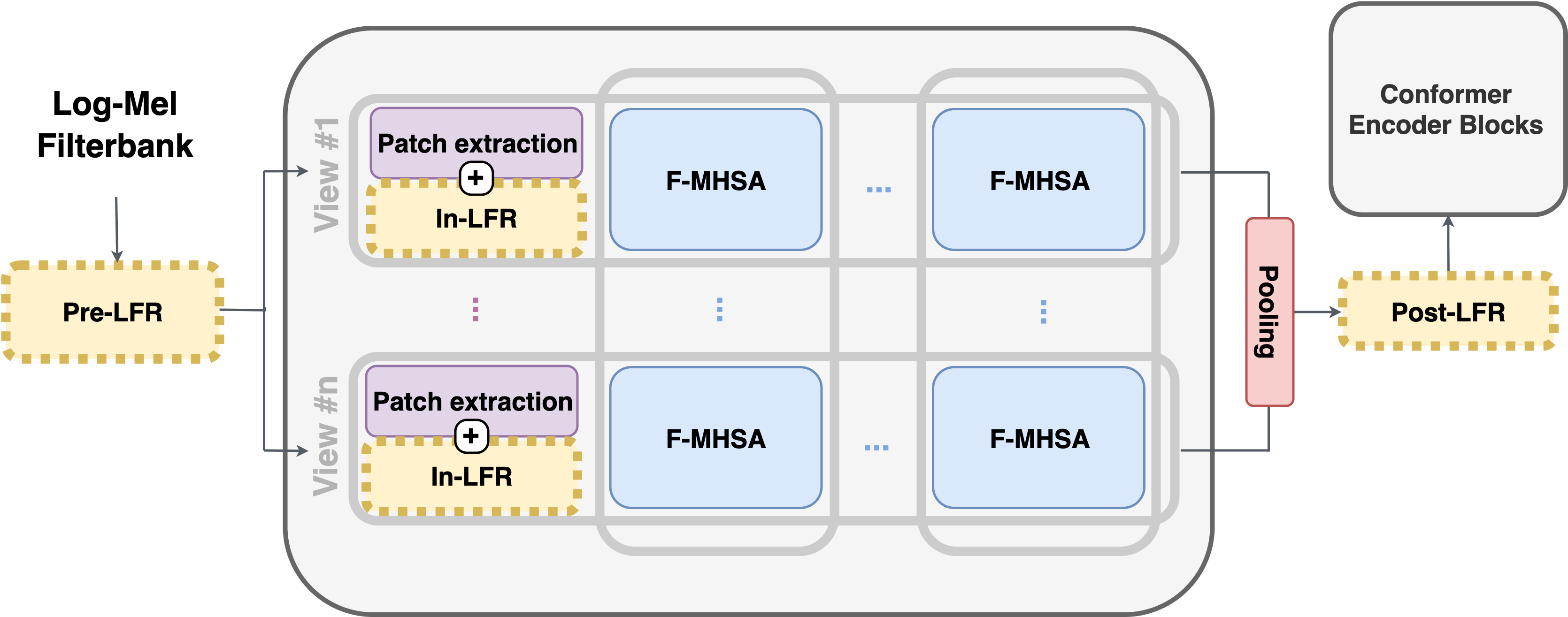}
    \caption{mvF-Attention frontend with configurable number of views and layers. The downsampling positions (Pre-LFR, In-LFR, Post-LFR) are optional and compared in Section~\ref{sec:lfr}.}
    \label{fig:mvfattention}
\end{figure}

The proposed frontend has two main steps: (1) patch extraction, with one or multiple patch sizes ("views"), and (2) F-Attention, with one or multiple layers of self-attention as illustrated in Figures \ref{fig:fattention_frontend} and \ref{fig:mvfattention}.
The layers of each view operate independently and are only merged at the end via a pooling operation.

To let the model learn different relationships, we extract patches of the spectrogram for each view using different patch sizes.
In each view we use multi-head self-attention to analyse the interactions between patches that correspond to the same time range but to different frequency bins.
We do not share parameters neither across the self-attention blocks inside a view nor within a layer (in different views) to ensure that each self-attention analyses different patterns. Following the Transformer architecture \cite{transformer}, each \gls{MHSA} is followed by a residual connection and layer normalization.

To merge the outputs of the different views, \glspl{mvfLSTM} concatenate the output of each view and use a linear layer to project it to the dimension of the encoder. However, we notice that that leads to an escalation of the number of the parameters in the linear layer when stacking many views. To bypass this problem, we decide to merge the different views using a pooling layer in the patch embedding dimension. Therefore, the size of the output of the frontend is kept constant regardless of the number of views, maintaining a reasonable number of parameters in the linear layer.

\section{Experimental Results}
This section describes the implementation details of our models, the datasets and training setup.
We conduct experiments on internal data with a conformer transducer architecture first and validate findings with a \gls{LAS} architecture on a public dataset subsequently.

\subsection{Voice-controlled far-field experiments}
We train a Conformer Transducer model \cite{gulati20_interspeech} with 73.5 M parameters on about 50 000 h of de-identified English speech from voice-controlled far-field devices.
In the Conformer Transducer the \gls{LSTM} encoder of an RNN-T model \cite{rnnt, rnnt2, rnnt3} is replaced with a Conformer encoder \cite{gulati20_interspeech}.
The baseline frontend, with a total of 3.3 M parameters, has 2 \gls{CNN} layers of kernel $3\times3$, stride $2\times2$ and an embedding dimension of 128. These are followed by a linear layer that projects the tensor to the encoder embedding dimension.
F-Attention frontends, with a number of parameters from 3.2 M to 3.5 M depending on the variant, have a patch embedding layer for each view, with different patch sizes and a stride of $4\times4$ so that the compression factor is equal to the baseline\footnote{Note that a patch size of $7\times7$ and a stride of $4\times4$ replicates the behaviour of the two convolutional layers in the baseline.}.
In all setups, we selected $7\times7$ patches when using one view and extended to $7\times7$, $14\times14$ and $3\times3$, $7\times7$, $14\times14$, $28\times28$ when using two and four views, respectively.
The patch embedding dimension is 128. Each of the self-attention layers in the frontend has 8 heads and a dimension of 128.
The Conformer encoder comprises 12 Conformer blocks with a total of 50.9 M parameters. Each Conformer block has 8 self-attention heads, and an encoder dimension of 512.
The predictor network consists of a $2\times768$ \gls{LSTM} with a total of 15.2 M parameters makes up the prediction network.
A feed-forward layer with 512 units and \textit{tanh} activation makes up the first layer of the joint network, which is followed by a \textit{softmax} layer with an output vocabulary size of 4 001 word pieces.

We train each model for 500 000 steps with the Adam optimizer \cite{adam}, a learning rate of 0.0001, a linear warm-up of 5 k iterations, and an exponential decay. Additionally, we group training examples by sequence length with bucket boundaries $\{300, 600\}$ and corresponding bucketing batch sizes of $[16, 8, 2]$.
Unless otherwise stated, the acoustic features are 64-dimensional \gls{lfbe} features with a window size of 25 ms and a frame shift of 10 ms. As an augmentation method, we  use a variant of SpecAugment \cite{park19e_interspeech}, as proposed in \cite{specaugment2}.
For the evaluation, the top 5 best checkpoints out of the last 10 are selected based on validation dataset \gls{WER}. Then, the weights of the top 5 models are averaged and used to obtain the final test scores.

Our experiments follow four main directions: (1) studying the position of \gls{lfr} conversion with respect to the proposed frontend, (2) analyzing the proposed frontend with variations in the number of views or attention layers, (3) studying the generalizability to different speech features, and (4) exploring the ability of the F-Attention to ignore noise and focus on relevant frequency ranges.

\subsubsection{Optimal position of \gls{lfr} conversion}
\label{sec:lfr}
In the baseline model, speech features are stacked and downsampled by a factor of 3, corresponding to an encoder frame rate of \SI{30}{ms}. Although this is done before the frontend, we hypothesize that F-Attention works optimally on less-processed features. Therefore, we consider three different approaches, shown in yellow in Figure \ref{fig:mvfattention}.
Pre-\gls{lfr} is the standard approach, features are stacked and downsampled before the frontend, as in the baseline. In-\gls{lfr} conversion consists of a modification of the strides and and patch shapes in the patch extraction layer so that the downsampling is done simultaneously. Finally, Post-\gls{lfr} conversion consists of the same transformation used in Pre-\gls{lfr} conversion, but it is employed after the frontend. For any of the three approaches, the final encoder frame rate is \SI{30}{ms}.

Table \ref{table:lfr} summarizes this initial study.
For a F-Attention frontend with one view and one layer we observe that Pre-\gls{lfr} as well as In-\gls{lfr} degrade over the baseline.
Only Post-\gls{lfr} is on par with the baseline.
Observing similar trends on the development partition we use the Post-\gls{lfr} setting going forward and do not revisit this decision for more views or more layers.
\begin{table}[h]
\centering
\begin{tabular}{llS[table-format=-2.1,table-text-alignment=center]}
\toprule
Frontend & LFR Position & {rWERR (\%, $\uparrow$)}\\
\midrule
Baseline & Pre-LFR & \bfseries 0.0 \\
F-Attention & Pre-LFR & -1.2 \\
F-Attention & In-LFR & -0.5 \\
F-Attention & Post-LFR & \bfseries 0.0 \\
\bottomrule
\end{tabular}
\caption{Study of different \gls{lfr} conversion strategies combined with our frontend in terms of \gls{rWERR}. $\uparrow$ : higher is better.}
\label{table:lfr}
\end{table}%
\subsubsection{Optimal configuration of layers and views}
To evaluate the performance of our frontend and find the optimal configuration, we explore different combinations on the number of views and layers, keeping the number of parameters close to the baseline.
While the baseline model has 73.5 M of which 3.3 M are the frontend, all our proposed frontends have between 3.2 M and 3.5 M parameters.
To study the consistency of our results, we train two variants of the models for each views/layers combination: Conformer transducer with joint network and without it (with a single linear projection).

The results shown in Table \ref{table:mvfattention} indicate consistent gains when using the F-Attention frontend with or without the joint network. Comparing the experiments with one layer and multiple views against those with one view but multiple layers, we can conclude that adding views can have a better impact on the model performance. Furthermore, observing the experiment with two views and two layers, we obtain the best performance obtained by any of our models. However, we decide not to try more configurations with simultaneously multiple views and layers due to the additional increase in the number of parameters and GPU memory requirements.
\begin{table}[t]
\centering
\begin{tabular}{
c
S[table-format=1.1,table-text-alignment=center]
S[table-format=-2.1,table-text-alignment=center,retain-explicit-plus,input-symbols={*}]
S[table-format=-2.1,table-text-alignment=center,retain-explicit-plus,input-symbols={*}]
}
\toprule
\multirow{2}[2]{*}{\begin{tabular}{c}Layers/\\views\end{tabular}} &
{\multirow{2}[2]{*}{\begin{tabular}{c}Params\\(M)\end{tabular}}} &
\multicolumn{2}{c}{rWERR(\%, $\uparrow$)} \\
\cmidrule{3-4}
& & {w/o Joint Net} & {w/ Joint Net} \\
\midrule
Baseline & 3.3 & 0.0{*}  &  0.0{**} \\
 1 / 1  &  3.2 & 0.0  & -0.4 \\ 
 1 / 2  &  3.3 & \bfseries +2.1 & +1.3 \\
 1 / 4  & 3.5  & +1.9 & +2.2 \\ 
 2 / 1  &  3.3 & +1.4 & +1.1 \\ 
 4 / 1  & 3.4  & +0.9 & +0.7 \\ 
 2 / 2  & 3.4  & +1.2 & \bfseries +2.4 \\
\bottomrule
\end{tabular}
\caption{Study of different number of layers and views of the F-Attention frontend. For every experiment we compare the number of parameters of the frontend and the performance of the model measured on \gls{rWERR}. Each model is compared to the respective baseline with or without joint network. $\uparrow$ : higher is better. *: Corresponds to last row in Table \ref{table:lfr}. **: +5.8 wrt. baseline without joint network.}
\label{table:mvfattention}
\end{table}

\subsubsection{Behaviour on Different Speech Features}
To understand how well the F-Attention frontend generalizes to other features, we perform a brief comparison of F-Attention frontends (choosing the two more comparable in number of parameters) on 256-dimensional \gls{lstft} features instead of 64-dimensional \gls{lfbe} features used in previous experiments.

We observe that a model with F-Attention frontend with 1 layer and 2 views improves by \SI{1}{\percent} relative over the corresponding baseline with convolutional frontend whereas an F-Attention frontend with 2 layers and 1 view just leads to \SI{0.4}{\percent} \gls{rWERR}.
We conclude that the approach generalizes to this feature variant but concentrate on \gls{lfbe} going forward.

\subsection{Interpretability Analysis of the mvF-Attention}
\label{ssec:interpretability}
The use of F-Attention enables the model to focus on specific frequency ranges and dismiss the information on others. This ability can be useful on noisy backgrounds, where the frontend could be able to ignore the noisiest frequency ranges. 

\begin{figure}[ht]
    \centering
    \includegraphics[scale=.125]{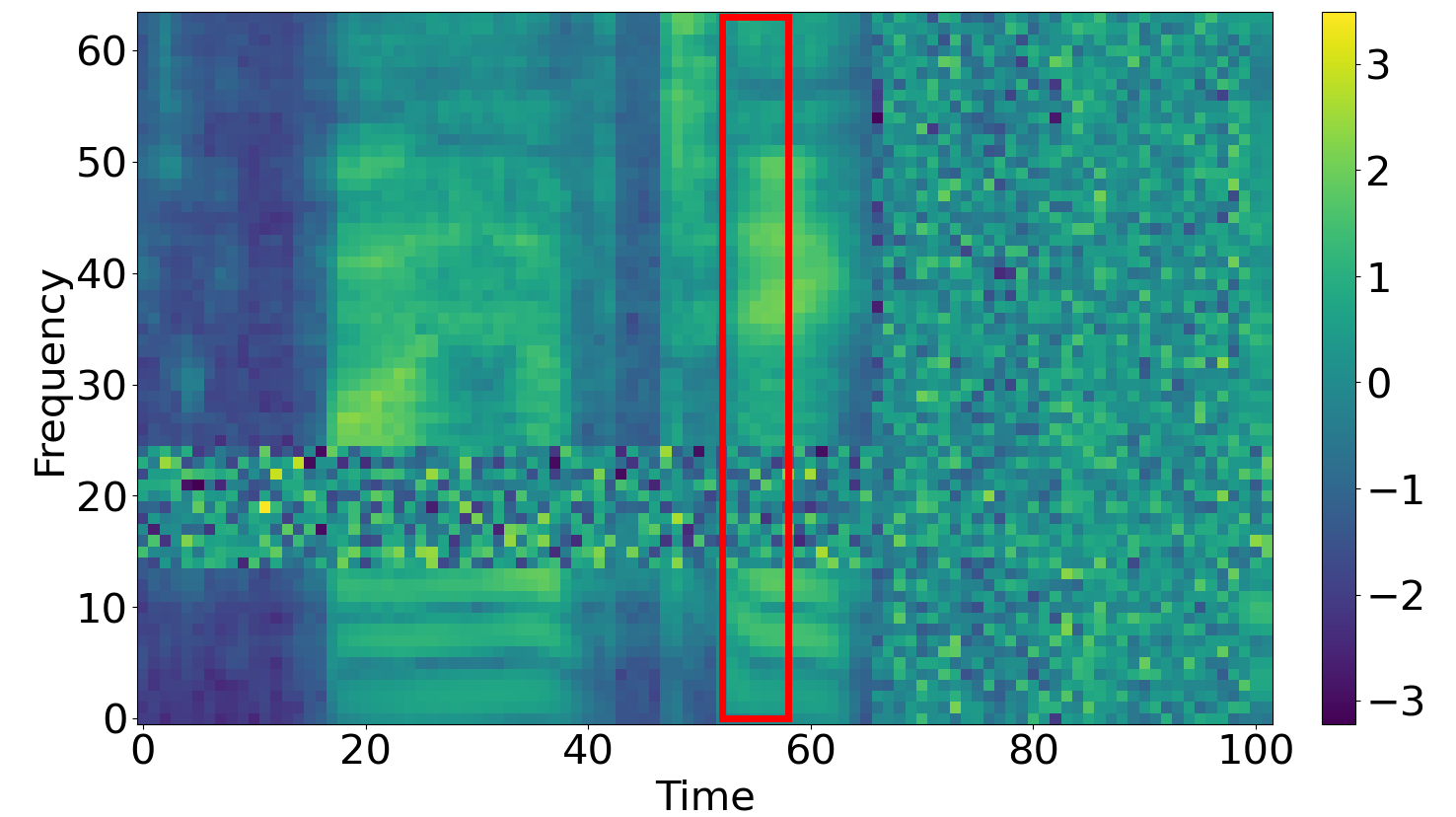}
    \caption{Analysis of a spectrogram taken from the public LibriSpeech corpus. Time range to study highlighted red.}
    \label{fig:spectrogram}
\end{figure}

\begin{figure}[ht]
    \centering
    \includegraphics[scale=.125]{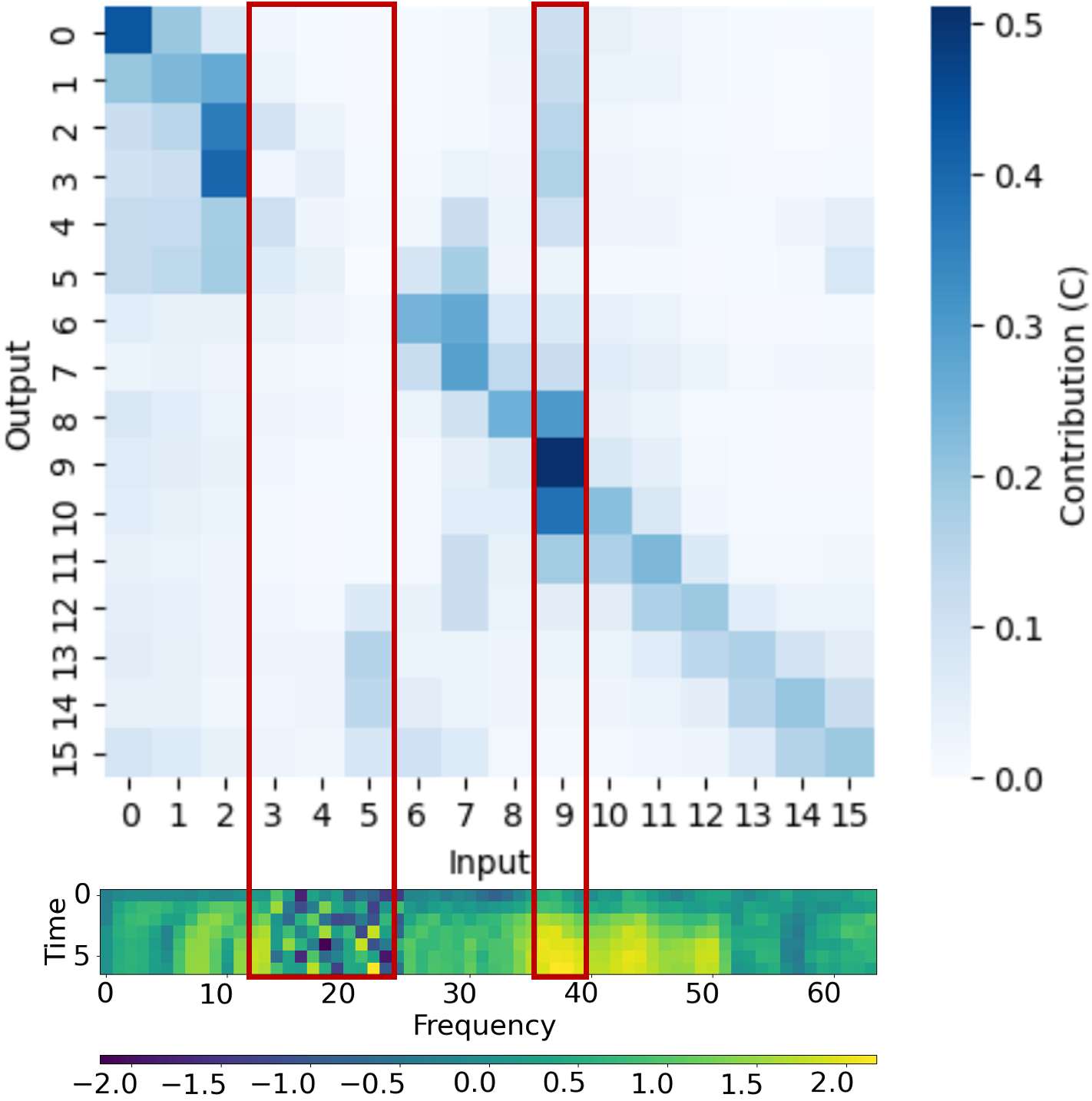}
    \caption{Zoom of the analyzed time range and token attributions. We see how patches that contain noise added by Spec Augment (from 3rd to 5th) don't contribute to any output token.}
    \label{fig:contributions}
\end{figure}

In this regard, aiming to understand if F-Attention is processing the spectrogram as expected, we perform an interpretability analysis on the frontend. To do so, we employ ALTI \cite{ferrando-2022-measuring}, a token attribution method that summarizes information from the self-attention layer (attention weights, values and output projection), the residual connection and the layer normalization. Using this method, we obtain a contributions plot (Figure \ref{fig:contributions}, $C$ is defined as in Eq. 10 in \cite{ferrando-2022-measuring}) that shows how every input token (columns) contributes to each output token (rows) when doing a forward pass of the embedded patches through the F-Attention layer. For the sake of simplicity, we decide to analyse our most simple model, the F-Attention frontend with one view and one layer. The F-Attention layer operates inside specific time ranges, extracting interactions frequency-wise. We analyse the behaviour of the frontend on a specific time range shown in Figure \ref{fig:spectrogram}.  

Analysing the results in Figure \ref{fig:contributions}, we can see how those input tokens that correspond to patches that contain noise added by SpecAugment (from 3rd to 5th) have a negligible contribution to every output token. Additionally, we see that patch number 9, which represents a highlighted frequency range has a big impact to many output tokens. Therefore, the F-Attention layer is able to detect the most relevant frequency ranges and ignores those that are not useful for the transcription.

\subsection{Verification on public data}
To verify the efficacy of the proposed front-end on public data, we also evaluate the approach on LibriSpeech. For this dataset, the model consists of six-layer bi-directional \gls{LSTM} \cite{10.1162/neco.1997.9.8.1735} with 1,024 units encoder and a two-layer uni-directional \gls{LSTM} decoder with a single head content-based attention mechanism~\cite{Chorowski2015AttentionBasedMF}. Overall the model has 104M trainable parameters. The model has been trained with Adam optimizer~\cite{Kingma2015AdamAM}. We use a SentencePiece~\cite{kudo2018sentencepiece} unigram word-piece model with 4000 tokens. SpecAugment data augmentation method with LD policy~\cite{Park2019SpecAugmentAS} was used throughout model training. The model was trained with weight noise~\cite{Graves2011PracticalVI}, which was added to all encoder trainable parameters by sampling noise from a normal distribution with a standard deviation of 0.075 starting from 15k training steps. We use uniform label smoothing~\cite{Chorowski2017TowardsBD} which distributes a probability mass of 0.1 to non-ground truth tokens.

Table~\ref{table:librispeech} summarizes results for different combinations of number of layers, number of views and patch embedding size $E$.
In line with aforementioned joint network results on internal data the configuration with two layers and two views performs best, leading to \SI{4.6}{\percent} \gls{rWERR} on the \emph{test other} partition.
We further observe that reducing the patch embedding size $E$ reduces the number of parameters but shows mixed results when comparing the \emph{test clean} and the \emph{test other} partition.
In all cases, adding more views leads to better performance on \emph{test other} with mixed results on \emph{test clean}.

\begin{table}[t]
\centering
\begin{tabular}{@{}
c@{}
S[table-format=3,table-text-alignment=center]@{}
S[table-format=3.1,table-text-alignment=center]@{}
S[table-format=1.1,table-text-alignment=center]@{}
S[table-format=1.1,table-text-alignment=center]
S[table-format=-2.1,table-text-alignment=center,retain-explicit-plus]@{}
S[table-format=-2.1,table-text-alignment=center,retain-explicit-plus]@{}
}
\toprule
{\multirow{2}[2]{*}{\begin{tabular}{c}Layers/\\views\end{tabular}}} &
{\multirow{2}[2]{*}{$E$}} &
{\multirow{2}[2]{*}{\begin{tabular}{c}Params\\(M)\end{tabular}}} &
\multicolumn{2}{c}{WER(\%, $\downarrow$)} & 
\multicolumn{2}{c}{rWERR(\%, $\uparrow$)} \\
\cmidrule(r){4-5}
\cmidrule(l){6-7}
& & & {Clean} & {Other} & {Clean} & {Other} \\
\midrule
Baseline &  & 104.7 & 3.1 & 8.2 & +0.0 & +0.0 \\
1 / 1 & 128 & 105.8 & 3.1 & 8.1 & -0.6 & +1.3 \\
2 / 1 & 128 & 105.9 & 3.1 & 8.1 & -0.1 & +1.0 \\
2 / 2 & 128 & 106.0 & \bfseries3.0 & \bfseries 7.8 & \bfseries +3.9 & \bfseries +4.6 \\
1 / 1 & 64 & 105.2 & 3.1 & 8.3 & +0.4 & -1.6 \\
1 / 2 & 64 & 105.3 & \bfseries 3.0 & 8.2 & +1.7 & -0.5 \\
1 / 4 & 64 & 105.3 & 3.1 & 8.2 & -1.7 & +0.1 \\
\bottomrule
\end{tabular}
\caption{LibriSpeech test results. Comparison of F-Attention frontends against a convolutional frontend using an \gls{LSTM}-based \gls{LAS} model with a patch embedding size $E$.}
\label{table:librispeech}
\end{table}

\subsubsection{Robustness to Noise Addition}
\begin{figure}[ht]
    \centering
    \includegraphics[scale=.16]{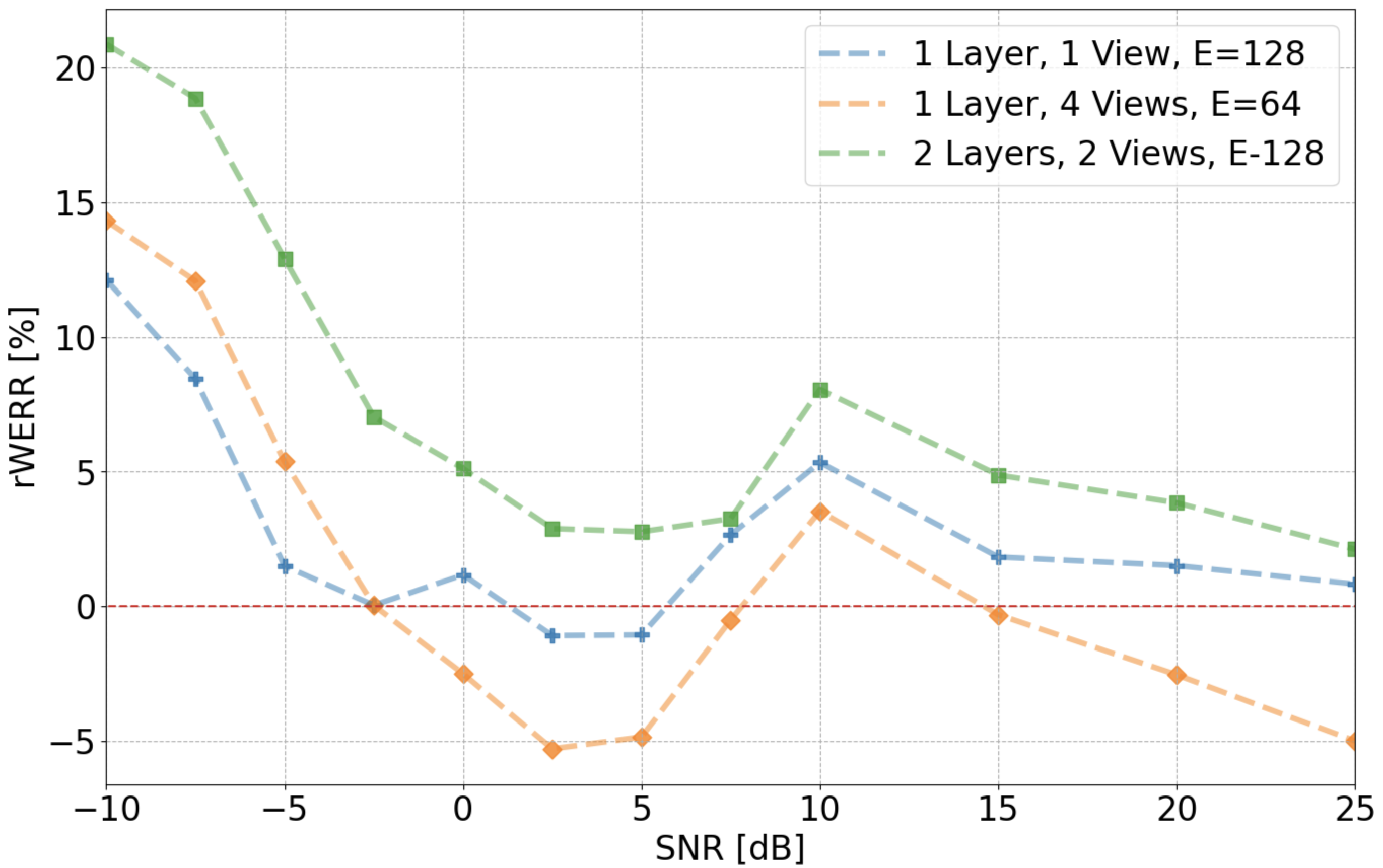}
    \caption{Robustness of different front-ends in comparison to the baseline for different \gls{SNR} conditions on LibriSpeech with internally-recorded bubble background noise. All results are \gls{rWERR}. Higher is better.}
    \label{fig:snr}
\end{figure}

The results shown in section \ref{ssec:interpretability} show that F-Attention is able to ignore noise and focus on relevant frequency ranges.
To extend this study we evaluate our best LibriSpeech models in terms of performance and comparability to the baseline on the same test set but with different noise conditions.
We decided to add internally-recorded babble background noise to utterances of the test clean partition of LibriSpeech.
Figure \ref{fig:snr} shows \gls{rWERR} for different different \gls{SNR} conditions.
We observe that the evaluated configurations improve substantially over the baseline in very noisy conditions, i.e. $\text{SNR} < 0$.
This effect is most pronounced for the 1-layer 1-view F-attention configuration.
Interestingly, all systems improve over the baseline at about \SI{10}{dB} \gls{SNR}.
A possible explanation is, that a certain noise floor covers up artifacts of the processing pipeline or that slightly noisy conditions better reflect the training conditions.

\section{Conclusions}
In this paper we have proposed frequency attention as a new Transformer frontend that processes speech using self-attention over frequency bins.
Our proposed frontend leads to \SI{2.4}{\percent} \gls{rWERR} compared to a convolutional frontend on large in-house data with a Conformer-Transducer model.
We demonstrated that these gains transfer to public data (here LibriSpeech) and a different \gls{asr} architecture (here \gls{LAS}).
On the LibriSpeech \emph{test other} partition, the proposed frontend obtains a \SI{4.6}{\percent} \gls{rWERR}.
We demonstrated that these findings hold for \gls{lstft} and \gls{lfbe} features. Using an interpretability method we have been able to see how the frontend discards noise and focuses on relevant frequency ranges. This insight was further corroborated by a analyzing the F-attention performance in a wide range of \glspl{SNR} with relative gains over \SI{20}{\percent} in very noisy conditions.

\newpage
\bibliographystyle{IEEEtran}
\bibliography{mybib}

\begin{thebibliography}{10}
\providecommand{\url}[1]{#1}
\csname url@samestyle\endcsname
\providecommand{\newblock}{\relax}
\providecommand{\bibinfo}[2]{#2}
\providecommand{\BIBentrySTDinterwordspacing}{\spaceskip=0pt\relax}
\providecommand{\BIBentryALTinterwordstretchfactor}{4}
\providecommand{\BIBentryALTinterwordspacing}{\spaceskip=\fontdimen2\font plus
\BIBentryALTinterwordstretchfactor\fontdimen3\font minus
  \fontdimen4\font\relax}
\providecommand{\BIBforeignlanguage}[2]{{%
\expandafter\ifx\csname l@#1\endcsname\relax
\typeout{** WARNING: IEEEtran.bst: No hyphenation pattern has been}%
\typeout{** loaded for the language `#1'. Using the pattern for}%
\typeout{** the default language instead.}%
\else
\language=\csname l@#1\endcsname
\fi
#2}}
\providecommand{\BIBdecl}{\relax}
\BIBdecl

\bibitem{transformer}
A.~Vaswani, N.~Shazeer, N.~Parmar, J.~Uszkoreit, L.~Jones, A.~N. Gomez,
  L.~Kaiser, and I.~Polosukhin, ``Attention is all you need,'' in
  \emph{Proceedings of the 31st International Conference on Neural Information
  Processing Systems}, ser. NIPS'17.\hskip 1em plus 0.5em minus 0.4em\relax Red
  Hook, NY, USA: Curran Associates Inc., 2017, p. 6000–6010.

\bibitem{dong_icassp}
L.~Dong, S.~Xu, and B.~Xu, ``Speech-transformer: A no-recurrence
  sequence-to-sequence model for speech recognition,'' in \emph{2018 IEEE
  International Conference on Acoustics, Speech and Signal Processing
  (ICASSP)}, 2018, pp. 5884--5888.

\bibitem{gulati20_interspeech}
A.~Gulati, J.~Qin, C.-C. Chiu, N.~Parmar, Y.~Zhang, J.~Yu, W.~Han, S.~Wang,
  Z.~Zhang, Y.~Wu, and R.~Pang, ``{Conformer: Convolution-augmented Transformer
  for Speech Recognition},'' in \emph{Proc. Interspeech 2020}, 2020, pp.
  5036--5040.

\bibitem{gong21b_interspeech}
Y.~Gong, Y.-A. Chung, and J.~Glass, ``{AST: Audio Spectrogram Transformer},''
  in \emph{Proc. Interspeech 2021}, 2021, pp. 571--575.

\bibitem{wang-etal-2020-fairseq}
\BIBentryALTinterwordspacing
C.~Wang, Y.~Tang, X.~Ma, A.~Wu, D.~Okhonko, and J.~Pino, ``Fairseq {S}2{T}:
  Fast speech-to-text modeling with fairseq,'' in \emph{Proceedings of the 1st
  Conference of the Asia-Pacific Chapter of the Association for Computational
  Linguistics and the 10th International Joint Conference on Natural Language
  Processing: System Demonstrations}.\hskip 1em plus 0.5em minus 0.4em\relax
  Suzhou, China: Association for Computational Linguistics, Dec. 2020, pp.
  33--39. [Online]. Available: \url{https://aclanthology.org/2020.aacl-demo.6}
\BIBentrySTDinterwordspacing

\bibitem{cnns_lecun}
Y.~Lecun, L.~Bottou, Y.~Bengio, and P.~Haffner, ``Gradient-based learning
  applied to document recognition,'' \emph{Proceedings of the IEEE}, vol.~86,
  no.~11, pp. 2278--2324, 1998.

\bibitem{mvflstm}
\BIBentryALTinterwordspacing
M.~Van~Segbroeck, H.~Mallidih, B.~King, I.-F. Chen, G.~Chadha, and R.~Maas,
  ``Multi-view frequency lstm: An efficient frontend for automatic speech
  recognition,'' 2020. [Online]. Available:
  \url{https://arxiv.org/abs/2007.00131}
\BIBentrySTDinterwordspacing

\bibitem{dosovitskiy2021an}
\BIBentryALTinterwordspacing
A.~Dosovitskiy, L.~Beyer, A.~Kolesnikov, D.~Weissenborn, X.~Zhai,
  T.~Unterthiner, M.~Dehghani, M.~Minderer, G.~Heigold, S.~Gelly, J.~Uszkoreit,
  and N.~Houlsby, ``An image is worth 16x16 words: Transformers for image
  recognition at scale,'' in \emph{International Conference on Learning
  Representations}, 2021. [Online]. Available:
  \url{https://openreview.net/forum?id=YicbFdNTTy}
\BIBentrySTDinterwordspacing

\bibitem{devlin-etal-2019-bert}
\BIBentryALTinterwordspacing
J.~Devlin, M.-W. Chang, K.~Lee, and K.~Toutanova, ``{BERT}: Pre-training of
  deep bidirectional transformers for language understanding,'' in
  \emph{Proceedings of the 2019 Conference of the North {A}merican Chapter of
  the Association for Computational Linguistics: Human Language Technologies,
  Volume 1 (Long and Short Papers)}.\hskip 1em plus 0.5em minus 0.4em\relax
  Minneapolis, Minnesota: Association for Computational Linguistics, Jun. 2019,
  pp. 4171--4186. [Online]. Available: \url{https://aclanthology.org/N19-1423}
\BIBentrySTDinterwordspacing

\bibitem{rnnt}
A.~Graves, ``{Sequence transduction with recurrent neural networks},'' in
  \emph{Proceedings Representation Learning Workshop on International
  Conference on Machine Learning (ICML)}, Edinburgh, Scotland, 2012.

\bibitem{rnnt2}
C.~Zhang, B.~Li, Z.~Lu, T.~N. Sainath, and S.-y. Chang, ``Improving the fusion
  of acoustic and text representations in rnn-t,'' in \emph{ICASSP 2022 - 2022
  IEEE International Conference on Acoustics, Speech and Signal Processing
  (ICASSP)}, 2022, pp. 8117--8121.

\bibitem{rnnt3}
J.~Li, R.~Zhao, H.~Hu, and Y.~Gong, ``Improving rnn transducer modeling for
  end-to-end speech recognition,'' in \emph{2019 IEEE Automatic Speech
  Recognition and Understanding Workshop (ASRU)}, 2019, pp. 114--121.

\bibitem{adam}
\BIBentryALTinterwordspacing
D.~P. Kingma and J.~Ba, ``Adam: {A} method for stochastic optimization,'' in
  \emph{3rd International Conference on Learning Representations, {ICLR} 2015,
  San Diego, CA, USA, May 7-9, 2015, Conference Track Proceedings}, Y.~Bengio
  and Y.~LeCun, Eds., 2015. [Online]. Available:
  \url{http://arxiv.org/abs/1412.6980}
\BIBentrySTDinterwordspacing

\bibitem{park19e_interspeech}
D.~S. Park, W.~Chan, Y.~Zhang, C.-C. Chiu, B.~Zoph, E.~D. Cubuk, and Q.~V. Le,
  ``{SpecAugment: A Simple Data Augmentation Method for Automatic Speech
  Recognition},'' in \emph{Proc. Interspeech 2019}, 2019, pp. 2613--2617.

\bibitem{specaugment2}
D.~S. Park, Y.~Zhang, C.-C. Chiu, Y.~Chen, B.~Li, W.~Chan, Q.~V. Le, and Y.~Wu,
  ``Specaugment on large scale datasets,'' in \emph{ICASSP 2020 - 2020 IEEE
  International Conference on Acoustics, Speech and Signal Processing
  (ICASSP)}, 2020, pp. 6879--6883.

\bibitem{ferrando-2022-measuring}
\BIBentryALTinterwordspacing
J.~Ferrando, G.~I. G{\`a}llego, and M.~R. Costa-juss{\`a}, ``Measuring the
  mixing of contextual information in the transformer,'' in \emph{Findings of
  the Association for Computational Linguistics: EMNLP 2022}.\hskip 1em plus
  0.5em minus 0.4em\relax Abu Dhabi, United Arab Emirates: Association for
  Computational Linguistics, Dec. 2022, p. 8698–8714. [Online]. Available:
  \url{https://aclanthology.org/2022.emnlp-main.595}
\BIBentrySTDinterwordspacing

\bibitem{10.1162/neco.1997.9.8.1735}
\BIBentryALTinterwordspacing
S.~Hochreiter and J.~Schmidhuber, ``Long short-term memory,'' \emph{Neural
  Comput.}, vol.~9, no.~8, p. 1735–1780, Nov. 1997. [Online]. Available:
  \url{https://doi.org/10.1162/neco.1997.9.8.1735}
\BIBentrySTDinterwordspacing

\bibitem{Chorowski2015AttentionBasedMF}
J.~Chorowski, D.~Bahdanau, D.~Serdyuk, K.~Cho \emph{et~al.}, ``Attention-based
  models for speech recognition,'' in \emph{NIPS}, 2015.

\bibitem{Kingma2015AdamAM}
D.~P. Kingma and J.~Ba, ``Adam: A method for stochastic optimization,''
  \emph{CoRR}, vol. abs/1412.6980, 2015.

\bibitem{kudo2018sentencepiece}
T.~Kudo and J.~Richardson, ``Sentencepiece: A simple and language independent
  subword tokenizer and detokenizer for neural text processing,'' 2018.

\bibitem{Park2019SpecAugmentAS}
D.~S. Park, W.~Chan, Y.~Zhang, C.~Chiu \emph{et~al.}, ``Specaugment: A simple
  data augmentation method for automatic speech recognition,'' in
  \emph{INTERSPEECH}, 2019.

\bibitem{Graves2011PracticalVI}
A.~Graves, ``Practical variational inference for neural networks,'' in
  \emph{NIPS}, 2011.

\bibitem{Chorowski2017TowardsBD}
J.~Chorowski and N.~Jaitly, ``Towards better decoding and language model
  integration in sequence to sequence models,'' in \emph{INTERSPEECH}, 2017.

\end{thebibliography}
\end{document}